\documentclass[twocolumn,showpacs,preprintnumbers,amsmath,amssymb,superscriptaddress]{revtex4}
\usepackage{amssymb}
\usepackage{graphicx,color}
\usepackage{bm}

\begin{document}

\title{Gluon shadowing and hadron production in heavy-ion collisions at LHC}

\author{Wei-Tian Deng}
\affiliation{Department of Physics, Shandong University, Jinan 250100, China}
\affiliation{Frankfurt Institute for Advanced Studies (FIAS)
Ruth-Moufang-Strasse 1, D-60438 Frankfurt am Main, Germany}

\author{Xin-Nian Wang}
\affiliation{Institute of Particle Physics, Central China  Normal University, Wuhan 430079, China}
\affiliation{Nuclear Science Division, MS 70R0319,
Lawrence Berkeley National Laboratory, Berkeley, California 94720}

\author{Rong Xu}
\affiliation{Institute of Particle Physics, Central China Normal University, Wuhan 430079, China}

\begin{abstract}
The recently published first measurement of charged hadron multiplicity density at mid-rapidity 
$dN_{ch}/d\eta=1584 \pm 4 ({\rm stat.}) \pm 76 ({\rm sys.})$ in central $Pb+Pb$ collisions at 
$\sqrt{s} = 2.76$ TeV by the ALICE Experiment at LHC is in good agreement with the HIJING2.0 prediction
within the experimental errors and theoretical uncertainties. The new data point is used
to carry out a combined fit together with the RHIC data to reduce the uncertainty in the
gluon shadowing parameter $s_{g}$ which controls the overall magnitude of 
gluon shadowing at small fractional momentum $x$ in HIJING2.0 model. Predictions
on the centrality dependence of charged hadron multiplicity density at mid-rapidity with reduced
uncertainties are given for $Pb+Pb$ collisions at $\sqrt{s}=2.76$ and 5.5 TeV. The centrality
dependence is surprisingly independent of the colliding energy similar to that in $Au+Au$ collisions
at RHIC  for most of  centralities starting at $N_{\rm part}=50$ (100) at $\sqrt{s}=2.76$ (7) TeV. 
However, it becomes stronger in peripheral collisions at higher colliding energies.

\end{abstract}


\maketitle

Bulk observables such as rapidity density of hadron multiplicity and transverse energy provide 
important information on the initial entropy and energy production in high-energy heavy-ion collisions. They
also provide constraints on the initial conditions for hydrodynamical study of the collective phenomenon
and hard probes such as jet propagation and suppression. Because of the non-perturbative and many-body
nature of the physics processes involved, a first principle calculation of the bulk hadron production is so far
unaccessible in Quantum Chromodynamics (QCD). Instead, one has to reply on theoretical and phenomenological
models to estimate the bulk hadron production in high-energy heavy-ion collisions. Even with constraints by
experimental data from the Relativistic Heavy-ion Colliders (RHIC) experiments \cite{Adler:2004zn}
there still exits large uncertainties in the theoretical and phenomenological estimates of charged
hadron multiplicities in heavy-ion collisions at the Large Hadron Collider (LHC) both among different
models and within each model \cite{Abreu:2007kv}.

Recently, ALICE Experiment at LHC published the first experimental data on the 
charged hadron multiplicity density at mid-rapidity in central $Pb+Pb$ collisions at 
$\sqrt{s} = 2.76$ TeV \cite{Aamodt:2010pb}. The measured $dN_{ch}/d\eta=1584 \pm 4 ({\rm stat.}) \pm 76 ({\rm sys.})$
for the top 5\% central $Pb+Pb$ collisions is larger than most of theoretical and phenomenological
predictions, including all the latest color-glass-condensate model calculations \cite{Aamodt:2010pb}.
Such an unexpected large hadron multiplicity will have important implications on the underlying mechanism for 
initial parton production. It will also have important consequences on the study of other phenomena
such as collective flow and jet quenching in $Pb+Pb$ collisions at the LHC energies since they
all depend on the initial condition for the bulk matter evolution.

The first ALICE data \cite{Aamodt:2010pb} is in good agreement with the HIJING2.0 prediction \cite{Li:2001xa,Deng:2010mv}
within the experimental errors and theoretical uncertainty which is controlled mainly by the
uncertainty in the parameterization of the unknown nuclear shadowing of gluon distribution.
The new HIJING parameterization of the gluon shadowing in nuclei \cite{Li:2001xa}
was constrained mainly by experimental data on charged hadron multiplicity and its energy and
centrality dependence in heavy-ion collisions at RHIC within HIJING2.0 \cite{Deng:2010mv} model
which is an updated version of the original HIJING1.0 model \cite{Wang:1991hta,Gyulassy:1994ew}.
At the LHC energies, initial parton production probes gluon distribution at much smaller fractional
momentum $x$. The range of the gluon shadowing parameter $s_{g}=0.17-0.22$ allowed by the RHIC data, 
which controls the overall  magnitude of the gluon shadowing at small $x$ in the new HIJING parametrization,  
leads to much larger uncertainties in the final charged hadron multiplicity, up to about 15\% in the
most central $Pb+Pb$ collisions at $\sqrt{s}=2.76$ TeV \cite{Deng:2010mv}. Within the same HIJING2.0 
model, the ALICE data provides a much stringent constraint on the gluon shadowing. In this note, we will carry
out a global fit of the combined RHIC data on the centrality dependence of charged hadron 
multiplicity in $Au+Au$ collisions and the new ALICE data in the most central $Pb+Pb$ collisions  
at LHC to provide a better constraint on the gluon shadowing parameter $s_{g}$. With a smaller
range of the gluon shadowing parameter $s_{g}=0.20-0.23$, we will predict with reduced uncertainty the centrality 
dependence of charged hadron multiplicity density in mid-rapidity for $Pb+Pb$ collisions at 
both $\sqrt{s}=2.76$ and 5.5 TeV.

HIJING \cite{Wang:1991hta,Gyulassy:1994ew} is essentially a two-component model for hadron 
production in high energy nucleon  \cite{Sjostrand:1987su,Chen:1988bx, Wang:1990qp} and nuclear
 collisions \cite{Blaizot:1987nc, Kajantie:1987pd}.  In this two-component model, one divides nucleon 
 interaction into soft and hard part separated by a cut-off $p_0$ in the transverse momentum
 transfer between colliding partons.  Jet production with transverse momentum $p_T>p_0$
 can be calculated within the collinear factorized perturbative QCD (pQCD) parton model while
the soft interaction is characterized by a parameter in the effective cross section $\sigma_{soft}$.
 These two parameters, $\sigma_{soft}$ and $p_0$ in HIJING,  are 
determined phenomenologically by fitting the experimental data on total cross sections and 
hadron multiplicity in $p+p/\bar{p}$ collisions \cite{Wang:1991us}.  HIJING2.0 \cite{Deng:2010mv}
is an updated version in which old  Duke-Own (DO) parameterization \cite{Duke:1983gd}  of parton
distribution functions (PDF's)  is replaced by the Gluck-Reya-Vogt(GRV) parameterization \cite{Gluck:1994uf}.
Because of the much larger gluon distribution in GRV than DO parameterization at small $x$, one has 
to assume an energy-dependent  cut-off $p_0(\sqrt{s})$ and soft cross 
section $\sigma_{soft}(\sqrt{s})$ \cite{Deng:2010mv}  in order to fit the experimental values of the total
and inelastic cross sections of $p+p/(\bar p)$ collisions. The  values of $p_{0}$ and $\sigma_{soft}$
are further constrained by the energy-dependence of the central rapidity density of the charged hadron multiplicities.
This updated version HIJING2.0 can describe most of the features of hadron production in $pp$ collisions at
colliding energies up to 7 TeV at LHC \cite{Wang:1991us}.

For high-energy heavy-ion collisions, both nuclear modification of the
parton distribution functions and jet quenching in final state interaction have to be considered. 
Jet quenching in general suppresses high transverse momentum hadrons \cite{Wang:1991xy}. 
If we assume the effects of parton and hadron final state interactions on the total hadron multiplicity 
to be negligible \cite{Wang:2000bf,Eskola:2000xq,Lin:2000cx}, the only uncertainty
for hadron multiplicity density in $A+A$ collisions comes from  the nuclear modification of parton distribution
functions at small $x$. HIJING2.0 employes the factorized form of parton distributions in nuclei,
\begin{equation}
f_a^A(x,Q^2)=AR_a^A(x,Q^2)f_a^N(x,Q^2),
\label{eq:shdw}
\end{equation}
where $R_a^A(x,Q^2)$ is nuclear modification factor as given by the new HIJING
parameterization \cite{Li:2001xa},
\begin{eqnarray}
R_q^A(x,b)&=&1.0+1.19\log^{1/6}\!\!\!A\;(x^3-1.2x^2+0.21x) \nonumber \\
        &-&s_q(b)\;(A^{1/3}-1)^{0.6}(1-3.5\sqrt{x})\nonumber \\
        &\times& \exp(-x^2/0.01),
\label{eq:shq} \\
R_g^A(x,b)&=&1.0+1.19\log^{1/6}\!\!\!A\;(x^3-1.2x^2+0.21x) \nonumber \\
        &-&s_g(b)\;(A^{1/3}-1)^{0.6}(1-1.5x^{0.35})\nonumber \\
        &\times&\exp(-x^2/0.004),
\label{eq:shg}
\end{eqnarray}
for quarks and gluons, respectively. 
The impact-parameter dependence of the shadowing is implemented through the parameters,
\begin{equation}
s_a(b)=s_a\frac{5}{3}(1-b^2/R_A^2),
\label{eq-shadow2}
\end{equation}
where $R_A=1.12 A^{1/3}$ is the nuclear size. The form of the impact parameter 
dependence is chosen to give rise to the centrality dependence of the pseudorapidity multiplicity density per participant 
pair $2 dN_{ch}/d\eta/N_{\rm part}$. The impact-parameter averaging is done with a weight of the thickness function of the nucleus
with a hard-sphere nuclear distribution. 

\begin{figure}
  \centering
 \includegraphics[width=0.45\textwidth]{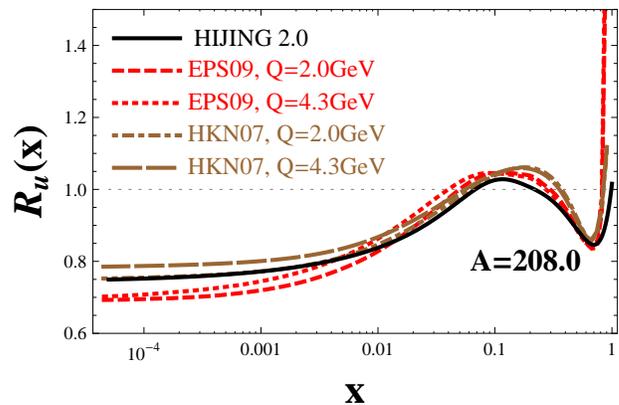}
  \caption{(color online) HIJING2.0 parameterization of nuclear shadowing factor for light quark distributions in lead nuclei (solid)
  with $s_{q}=0.1$ as compared to EPS09 (dashed and dotted)  \cite{Eskola:1998df} and HKN07 (dot-dashed and long dashed)  \cite{hkm}
  parameterizations at two different scales.}
    \label{fig:shq}
\end{figure}

The assumed form of nuclear modification in Eq.~(\ref{eq:shdw})
has been studied with data from deeply inelastic scattering (DIS) and Drell-Yan lepton pair production 
experiments \cite{Eskola:1998df, hkm}.  In the new HIJNG parametrization, the value $s_{q}=0.1$ is fixed by the 
experimental data on DIS off nuclear targets \cite{Li:2001xa}. Shown in Fig.~\ref{fig:shq} is the HIJING2.0 nuclear shadowing
factor for light quarks ($u$ and $d$) as compared to that by EPS09 \cite{Eskola:1998df} and HKN07 \cite{hkm} parameterizations
at two different scales $Q=2.0$ and 4.3 GeV/$c$. Both EPS09 and HKN07 use vacuum DGLAP evolution equations for the parton
distributions to determine the scale dependence of the nuclear shadowing factors which is rather weak for the quark
distributions. 

\begin{figure}
  \centering
 \includegraphics[width=0.45\textwidth]{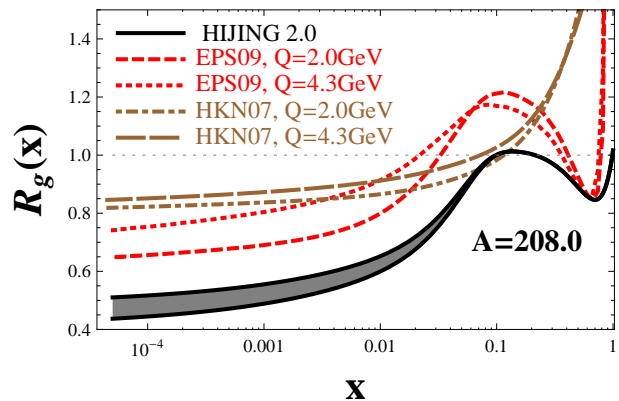}
  \caption{(color online) The same as Fig.~\ref{fig:shq} except for gluon distributions with $s_{g}=0.22-0.23$.}
    \label{fig:shg}
\end{figure}

The gluon shadowing at small $x$ is not constrained directly by DIS and Drell-Yan experimental data, except
momentum conservation. Total momentum depletion due to suppression of both quark and gluon distributions at small $x$
is partially compensated by the anti-shadowing at $x\approx 0.1$ for quarks. Momentum conservation, however, will not
provide much constraints on gluon distribution in nuclei both in the shadowing and anti-shadowing region as shown by EPS09
and HKN07 parameterizations in Fig.~\ref{fig:shg}, both enforce momentum conservation. HIJING2.0 parameterization violate
momentum sum rule by about 16\% which can be easily compensated by adjustment of gluon modification factor at large
and intermediate $x$ which will not affect mini-jet production at small $x$.

The value of gluon shadowing parameter $s_{g}$ in HIJING2.0 is constrained only by the hadron multiplicity in heavy-ion collisions. 
Using the combined RHIC data \cite{Adler:2004zn} on the centrality dependence of charged hadron multiplicity 
density in mid-rapidity as constraints, a range $s_{g}=0.17-0.22$ was obtained \cite{Deng:2010mv}. 
Note that HIJING2.0 parameterization assumes a scale-independent form. Such
an approximation for gluon distribution will not be valid at very large scale due to dominance of gluon emission dictated by the
DGLAP evolution equations. Both EPS09 and HKN07 parameterizations use the vacuum DGLAP evolution equations for parton
distributions to determine the scale dependence of the nuclear shadowing factors. At $Q=2.0$ and $4.3$ GeV/$c$, which are
the typical scales for minijet production at RHIC and LHC, respectively, the gluon shadowing varies at most about 13\% in EPS09
parameterization as shown in Fig.~\ref{fig:shg}. As shown in the figure, this is also approximately within the uncertainty in the HIJING2.0
parameterization of gluon shadowing constrained by the charged multiplicity data in heavy-ion collisions ($s_{g}=0.22-0.23$). 
Furthermore, higher-twist contributions to the DGLAP evolution equations might become important and would modify 
the scale dependence of the shadowing factor at intermediate scales.

\begin{figure}
  \centering
 \includegraphics[width=0.45\textwidth]{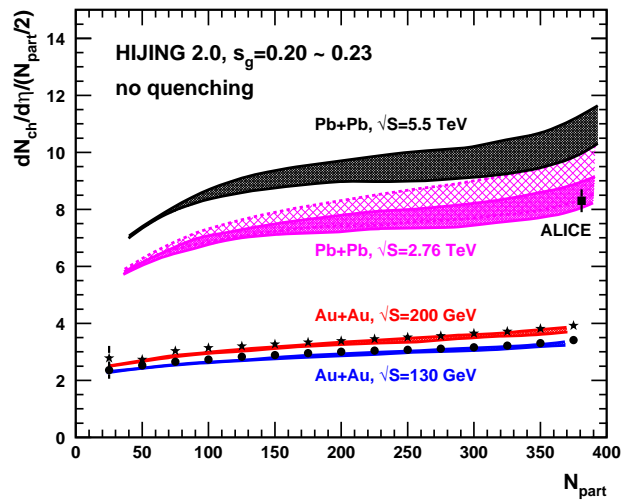}
  \caption{(color online) Charged hadron multiplicity density in mid-rapidity per participant pair
  $2dN_{ch}/d\eta/N_{\rm part}$ as a function of $N_{\rm part}$ from HIJING2.0 calculation
  with gluon shadowing parameter $s_{g}=0.20-0.23$ (solid-shade) and $s_{g}=0.17-0.22$ (dash-shade)
  as compared to combined RHIC data \cite{Adler:2004zn} for $Au+Au$ collisions (filled circle and star)
  and ALICE data \cite{Aamodt:2010pb} at LHC (solid square).}
  \label{fig:nch}
\end{figure}

With the above parameterization of 
parton shadowing and the range of gluon shadowing parameter $s_{g}=0.17-0.22$, the predicted
$2 dN_{ch}/d\eta/N_{\rm part}$, shown as dash-shaded area in Fig.~\ref{fig:nch}, agrees well
with the new ALICE data in the most central $Pb+Pb$ collisions at $\sqrt{s}=2.76$ TeV, within the
experimental error and  a large theoretical uncertainty of about 15\%  from the gluon 
shadowing parameter. The HIJING2.0 results are obtained by calculating $dN_{ch}/d\eta$ and $N_{\rm part}$
for different impact-parameters squared $b^{2}$ with equal bin size.
By performing a combined $\chi^{2}$-fit of the RHIC data \cite{Adler:2004zn} 
for $Au+Au$ collisions at $\sqrt{s}=200$ GeV and the data point from ALICE in the most 
central $Pb+Pb$ collisions at $\sqrt{s}=2.76$ TeV, the range of gluon shadowing parameter 
is reduced to $s_{g}=0.20-0.23$. As shown in Figs.~\ref{fig:shq} and \ref{fig:shg}, the HIJING2.0 gluon shadowing
from such fit is much stronger than the EPS09 and HKN07 parameterizations. However, it is comparable
to the parameterization by Strikman {\it et al.} \cite{strikman} and both are much stronger than the nuclear shadowing
for quark distributions.

With this new range of $s_{g}$ and therefore reduced uncertainty
we calculate the prediction for the centrality dependence of $dN_{ch}/d\eta$ in $Pb+Pb$ collisions
at both $\sqrt{s}=2.76$ and 5.5 TeV, shown in Fig.~~\ref{fig:nch}, as solid-shaded area. The 
calculated centrality dependence of $dN_{ch}/d\eta$ in $Au+Au$ collisions at two RHIC energies
is also shown together with combined RHIC data \cite{Adler:2004zn}.

\begin{figure}
  \centering
 \includegraphics[width=0.45\textwidth]{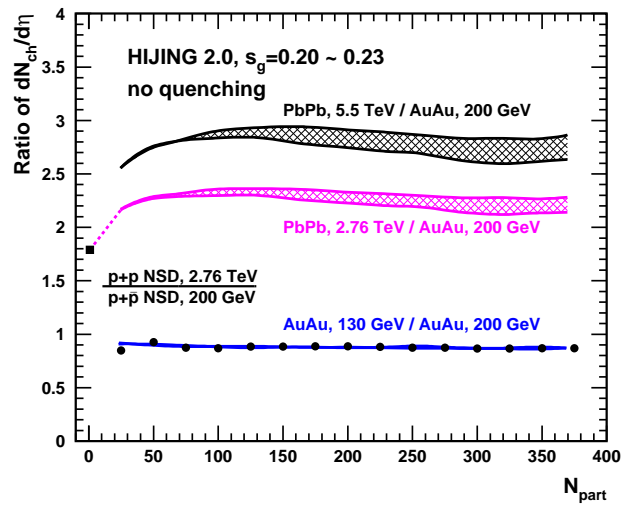}
  \caption{(color online) The ratio of charged hadron multiplicity density in mid-rapidity in
  heavy-ion collisions at different colliding energies, using $Au+Au$ collisions at $\sqrt{s}=200$ GeV
  as the common denominator. The data at the RHIC energies are combined from different
  experiments \cite{Adler:2004zn}. The ALICE data on non-single diffractive (NSD) $pp$ collisions 
  at $\sqrt{s}=2.36$ TEV  \cite{Aamodt:2010ft} is used to get the data point at $\sqrt{s}=2.76$ TeV
  using HIJING2.0 extrapolation. The NSD $p\bar p$ data at $\sqrt{s}=0.2$ TeV is from UA1 \cite{Albajar:1989an}}
  \label{fig:ratio}
\end{figure}

To exam the centrality dependence of $dN_{ch}/d\eta$ at different colliding energies in detail, 
we plot in Fig.~\ref{fig:ratio} the ratio of $dN_{ch}/d\eta$ at different colliding energies using 
$Au+Au$ collisions at $\sqrt{s}=0.2$ TeV as the common denominator.  In the figure we also plot
($pp$ 2.76 TeV)/($p\bar p$ 0.2 TeV), using the ALICE data on non-single diffractive $pp$ 
collisions at $\sqrt{s}=2.36 $ TeV \cite{Aamodt:2010ft} and HIJING2.0 calculation to extrapolate to the 
value at $\sqrt{s}=2.76$ TeV. The $p\bar p$ data at $\sqrt{s}=0.2$ TeV is from UA1 \cite{Albajar:1989an}.
The ratios of charged hadron multiplicity densities at the two LHC energies to that at RHIC are surprisingly flat over 
a large range of centralities just as the ratio of two RHIC energies.
It is interesting to note that the increased energy dependence of charged multiplicity density in
central $Pb+Pb$ collisions  over that in $pp$  is reached already at $N_{\rm part}=50 (100)$
for $\sqrt{s}=2.76$ (5.5) TeV. In other words, the centrality dependence of charged hadron
multiplicity density increases with energy in peripheral $Pb+Pb$ collisions.  Such centrality dependence 
of charged hadron is a consequence of the impact-parameter-dependent gluon shadowing in HIJING2.0. 

With a given transverse momentum cut-off $p_{0}$, the total number of
mini-jets per unit transverse area could become so large that it exceeds the limit
\begin{equation}
 \frac{T_{AA}(b)\sigma_{jet}}{\pi R_{A}^{2}}\le \frac{p_{0}^{2}}{\pi}
 \label{eq:limit}
\end{equation}
for independent multiple jet production for sufficiently large inclusive
jet cross section at high colliding energies and for large nuclei, where $T_{AA}(b)$ is the 
overlap function of $A+A$ collisions and $\pi/p_{0}^{2}$ is the intrinsic transverse size of 
a mini-jet with transverse momentum $p_{0}$. This is the reason for an energy-dependent 
cut-off $p_{0}$ for high-energy $pp$ collisions in HIJING2.0 since the GRV parton
distributions \cite{Gluck:1994uf} have a large gluon distribution at small $x$ and therefore
large mini-jet cross section at high colliding energies. The above limit for incoherent mini-jet
production should also depend on nuclear size and impact-parameter which can be
determined self-consistently through Eq.~(\ref{eq:limit}) \cite{Eskola:2000xq}. In HIJING2.0
such impact-parameter dependence of the cut-off scale is not considered. Instead, an
impact-parameter dependence of the gluon shadowing in Eq.~(\ref{eq-shadow2}) is considered
that is stronger than the typical nuclear length $L_{A}=\sqrt{R_{A}^{2}-b^{2}}$
dependence. Such a stronger impact-parameter dependence is favored by the centrality
dependence of $dN_{ch}/\eta$ in $Au+Au$ collisions at RHIC. This is also the reason for
nearly energy-independence of the centrality dependence of charged hadron multiplicity
density at the LHC energies. If confirmed by experimental measurements, it will have important implications 
on the initial eccentricity for the study of elliptic flow and jet quenching.

In summary, we have carried out a combined fit of the new ALICE data \cite{Aamodt:2010pb} on charged 
hadron multiplicity density in the most central $Pb+Pb$ collisions at $\sqrt{s}=2.76$ TeV and the
RHIC data within HIJING2.0 model. The range of gluon shadowing parameter $s_{g}=0.20-0.23$
in the new HIJING parameterization of parton shadowing \cite{Li:2001xa} enables us to predict
the centrality dependence of the charged hadron rapidity density with reduced uncertainty in
$Pb+Pb$ collisions at $\sqrt{s}=2.76$ and 5.5 TeV. The centrality dependence is surprisingly
independent of colliding energy for most centralities starting at $N_{\rm part}=50$ (100) for
$\sqrt{s}$=2.76 (5.5) TeV. However, the centrality dependence in the peripheral collisions
becomes stronger at higher colliding energies.

\section*{Acknowledgement}

This work was supported in part by the National Natural  Science Foundation of China under the project  
No. 10525523, No. 10825523, MOE of China under Project No. IRT0624, and the Director, Office of Energy
Research, Office of High Energy and Nuclear Physics, Division of Nuclear Physics, of the U.S. Department of 
Energy under Contract No. DE-AC02-05CH11231 and within the framework of the JET Collaboration. 
W.-T. Deng was also financially supported by Helmholtz
International Center for FAIR within the framework of the LOEWE program launched by the State of Hesse
during the completion of this work.


\begin{thebibliography}{88}


\bibitem{Adler:2004zn}
  S.~S.~Adler {\it et al.}  [PHENIX Collaboration],
  Phys.\ Rev.\  C {\bf 71}, 034908 (2005)
  [Erratum-ibid.\  C {\bf 71}, 049901 (2005)].


\bibitem{Abreu:2007kv}
  N.~Armesto {\it et al.},
  J.\ Phys.\ G {\bf 35}, 054001 (2008)
  [arXiv:0711.0974 [hep-ph]].

\bibitem{Aamodt:2010pb}
  K.~Aamodt {\it et al.}  [The ALICE Collaboration],
  arXiv:1011.3916 [nucl-ex].
  
  
  
\bibitem{Li:2001xa}
  S.~-Y.~Li, X.~-N.~Wang,
  Phys.\ Lett.\  {\bf B527}, 85-91 (2002).
  [nucl-th/0110075].

\bibitem{Deng:2010mv}
  W.~-T.~Deng, X.~-N.~Wang, R.~Xu,
  [arXiv:1008.1841 [hep-ph]].

  

\bibitem{Wang:1991hta}
  X.~N.~Wang and M.~Gyulassy,
  Phys.\ Rev.\  D {\bf 44}, 3501 (1991).

\bibitem{Gyulassy:1994ew}
  M.~Gyulassy and X.~N.~Wang,
  Comput.\ Phys.\ Commun.\  {\bf 83}, 307 (1994).
  
    
\bibitem{Sjostrand:1987su}
  T.~Sjostrand and M.~van Zijl,
  Phys.\ Rev.\  D {\bf 36}, 2019 (1987).
  T.~Sjostrand,
  Comput.\ Phys.\ Commun.\  {\bf 39}, 347 (1986).

  
\bibitem{Chen:1988bx}
  W.~R.~Chen and R.~C.~Hwa,
  Phys.\ Rev.\  D {\bf 39}, 179 (1989).

\bibitem{Wang:1990qp}
  X.~N.~Wang,
  Phys.\ Rev.\  D {\bf 43}, 104 (1991).


\bibitem{Blaizot:1987nc}
  J.~P.~Blaizot and A.~H.~Mueller,
  Nucl.\ Phys.\  B {\bf 289}, 847 (1987).

\bibitem{Kajantie:1987pd}
  K.~Kajantie, P.~V.~Landshoff and J.~Lindfors,
  Phys.\ Rev.\ Lett.\  {\bf 59}, 2527 (1987).
  K.~J.~Eskola, K.~Kajantie and J.~Lindfors,
  Nucl.\ Phys.\  B {\bf 323}, 37 (1989).


\bibitem{Wang:1991us}
  X.~-N.~Wang, M.~Gyulassy,
  Phys.\ Rev.\  {\bf D45}, 844-856 (1992).
  


\bibitem{Duke:1983gd}
  D.~W.~Duke and J.~F.~Owens,
  Phys.\ Rev.\  D {\bf 30}, 49 (1984).



\bibitem{Gluck:1994uf}
  M.~Gluck, E.~Reya and A.~Vogt,
  Z.\ Phys.\  C {\bf 67}, 433 (1995).

\bibitem{Wang:1991xy}
  X.~N.~Wang and M.~Gyulassy,
  Phys.\ Rev.\ Lett.\  {\bf 68}, 1480 (1992).

\bibitem{Wang:2000bf}
  X.~N.~Wang and M.~Gyulassy,
  Phys.\ Rev.\ Lett.\  {\bf 86}, 3496 (2001).

\bibitem{Eskola:2000xq}
  K.~J.~Eskola, K.~Kajantie and K.~Tuominen,
  Phys.\ Lett.\  B {\bf 497}, 39 (2001).

\bibitem{Lin:2000cx}
  Z.~W.~Lin, S.~Pal, C.~M.~Ko, B.~A.~Li and B.~Zhang,
  Phys.\ Rev.\  C {\bf 64}, 011902 (2001).


\bibitem{Eskola:1998df}
K.~J.~Eskola, H.~Paukkunen and C.~A.~Salgado,
  JHEP {\bf 0904}, 065 (2009)
  [arXiv:0902.4154 [hep-ph]].
  
 
  \bibitem{hkm}
M.~Hirai, S.~Kumano and M.~Miyama,
Phys.\ Rev.\ D {\bf 64}, 034003 (2001).

\bibitem{strikman}
  L.~Frankfurt, V.~Guzey and M.~Strikman,
  Phys.\ Rev.\  D {\bf 71}, 054001 (2005)
  [arXiv:hep-ph/0303022].
  
  

\bibitem{Aamodt:2010ft}
  K.~Aamodt {\it et al.} [ ALICE Collaboration ],
  Eur.\ Phys.\ J.\  {\bf C68}, 89-108 (2010).
  [arXiv:1004.3034 [hep-ex]].


\bibitem{Albajar:1989an}
 C.~Albajar {\it et al.}  [UA1 Collaboration],
 Nucl.\ Phys.\  B {\bf 335}, 261 (1990).




\end{thebibliography}
\end{document}